\documentclass[
twocolumn,
trackchanges
]{aastex631}

\usepackage{graphicx}
\usepackage{graphics} % for pdf, bitmapped graphics files
\usepackage{epsfig} % for postscript graphics files
\usepackage{mathptmx} % assumes new font selection scheme installed
\usepackage{times} % assumes new font selection scheme installed
\usepackage{amsmath} % assumes amsmath package installed
\usepackage{amssymb}  % assumes amsmath package installed
\usepackage{xcolor}
\usepackage{subfigure}
\usepackage{bm}

\submitjournal{\apjl}

\newcommand{\rs}{\text{R}_\odot}
\newcommand{\Mt}{\text{M}_{\text{t}}}

\begin{document}

\title{Compressible Turbulence in the Near-Sun Solar Wind: Parker Solar Probe's First Eight Perihelia}

\correspondingauthor{Manuel Enrique Cuesta}
\email{mecuesta@udel.edu}

\author[0000-0002-7341-2992]{Manuel Enrique Cuesta}
%\email{mecuesta@udel.edu}
\affiliation{Department of Physics and Astronomy, Bartol Research Institute, University of Delaware, Newark, DE 19716, USA}

\author[0000-0002-7174-6948]{Rohit Chhiber}
\affiliation{Department of Physics and Astronomy, Bartol Research Institute, University of Delaware, Newark, DE 19716, USA}
\affiliation{NASA Goddard Space Flight Center, Greenbelt, MD 20771, USA}
%\email{rohitc@udel.edu}

\author[0000-0002-4305-6624]{Xiangrong Fu}
\affiliation{New Mexico Consortium, Los Alamos, NM 87544}
\affiliation{Los Alamos National Laboratory, Los Alamos, NM 87545}

\author[0000-0003-1134-3909]{Senbei Du}
\affiliation{Los Alamos National Laboratory, Los Alamos, NM 87545}

\author[0000-0003-2965-7906]{Yan Yang}
\affiliation{Department of Physics and Astronomy, Bartol Research Institute, University of Delaware, Newark, DE 19716, USA}

\author[0000-0003-4168-590X]{Francesco Pecora}
\affiliation{Department of Physics and Astronomy, Bartol Research Institute, University of Delaware, Newark, DE 19716, USA}

\author[0000-0001-7224-6024]{William~H. Matthaeus}
%\email{whm@udel.edu}
\affiliation{Department of Physics and Astronomy, Bartol Research Institute, University of Delaware, Newark, DE 19716, USA}

\author[0000-0003-3556-6568]{Hui Li}
\affiliation{Los Alamos National Laboratory, Los Alamos, NM 87545}

\author[0000-0003-2491-1661]{John Steinberg}
\affiliation{Los Alamos National Laboratory, Los Alamos, NM 87545}

\author[0000-0003-4315-3755]{Fan Guo}
\affiliation{Los Alamos National Laboratory, Los Alamos, NM 87545}

\author[0000-0003-3886-0383]{Zhaoming Gan}
\affiliation{New Mexico Consortium, Los Alamos, NM 87544}
\affiliation{Los Alamos National Laboratory, Los Alamos, NM 87545}

\author{Emma Conrad}
\affiliation{Los Alamos National Laboratory, Los Alamos, NM 87545}

\author[0000-0001-7496-9832]{Diana Swanson}
\affiliation{University of New Hampshire, Durham NH 03824}
\affiliation{Los Alamos National Laboratory, Los Alamos, NM 87545}

\begin{abstract}

Many questions remain about the compressibility of solar wind turbulence 
with respect to its origins and properties.
Low plasma beta (ratio of thermal to magnetic pressure) environments allow for the easier generation of compressible turbulence, enabling 
study of the relationship between density fluctuations and turbulent Mach number.
Utilizing Parker Solar Probe plasma data, we examine the normalized proton density fluctuations \(\langle \delta n_p^2 \rangle ^{1/2}/\langle n_p\rangle = \delta {n_p}_{rms}/\langle n_p\rangle\) as a function of turbulent Mach number \(M_t\) conditioned on plasma beta and cross helicity.
With consideration of statistical error in the parameters computed from in-situ data, we find a general result that \(\delta {n_p}_{rms}/\langle n_p\rangle \sim M_t^{1.18 \pm 0.04}\), consistent with both linear-wave theory, and nearly-incompressible turbulence in an inhomogeneous background field.
We compare observational results conditioned on plasma beta and cross helicity with 3D magnetohydrodynamic simulations, and observe rather significant similarities with respect to how those parameters affect the proportionality between density fluctuations and turbulent Mach number.
This study further investigates the complexity of compressible turbulence as viewed by the density scaling relationship, and may help better understand the compressible environment of the near-Sun solar wind.

\end{abstract}


\section{Introduction}\label{sec:intro}

%\section{Background}\label{sec:background}

    Density fluctuations 
    persist on average at a low level of about 10\% in the solar wind inertial range
    and over a wide range of  heliocentric distances \citep{roberts1987JGRa}.
    The properties and origins 
    of this compressible component of turbulence are still unclear.
    Density fluctuations are generated from a non-zero divergence of the velocity field; however,
    when available, the regime of low plasma beta \(\beta\) (ratio of thermal to magnetic pressure) is a signature for %allows for 
    an easier generation of compressible turbulence \citep{roberts1987JGRa,roberts1987JGRb,roberts1990JGR,BavassanoBruno1995JGRDensity,GrappinEA1990, MalaraEAJGR1996}, a consequence of the plasma being dragged and compressed by the dominant magnetic pressure, warranting a more detailed investigation.
    %Therefore this subject warrants a more detailed investigation. 
    Such is the case in regions near recent Parker Solar Probe (PSP) perihelia due to the presence of a large (dominantly radial) magnetic field \citep{KasperEAprl2021} and the orbital approach to the coronal plasma, presumably of lower \(\beta\) \citep[e.g.,][]{chhiber2019psp2}. 
    
    The coexistence of compressive and incompressive fluctuations in magnetohydrodynamic (MHD) turbulence has been observed in the solar wind, for example, by comparing spectra of density and magnetic field fluctuations, which have a similar power-law nature \citep{MontgomeryEA1987JGR,KleinEA1993JGRDensity}.
    Theories involving small-amplitude and slow time scale compressive fluctuations have been used to expose the roles of Alfv\'en waves, pressure balances, and compressive waves in turbulent plasmas.
    Formally this involves an expansion of the compressible MHD equations about an incompressible state, based on small turbulent Mach numbers (\(M_t\)).
   This approach enables solutions characteristic of small-amplitude density fluctuations occurring at slow time scales \citep{matthaeus1988PoF,MatthaeusEA1991JGR}, namely nearly-incompressible (NI) theory.  
   For a homogeneous background field, NI theory predicts that the root-mean-square (\textit{rms}) density fluctuations scale with \(M_t^2\).
   This theory is derived most directly when plasma beta is large, i.e., sound speed greater than Alfv\'en speed ($c_s > V_A$) which is typically not the case for heliocentric distances \(R<0.30~{\rm au}\).  However large $V_A$ need not invalidate NI dynamics if the associated wave frequencies 
    ${\bf k} \cdot \bf{V}_A$ remain small ($kV_A\ll 1$) due to spectral (wavevector $\bf k$) anisotropy. 
      Indeed extensions of the theory to arbitrary plasma beta
    have also been presented, and these explicitly account for anisotropy relative to the mean magnetic field \citep{ZankEA1990GeoRLDensity,zank1992JPP,zank1993nearly, BaylyEA92}. These extensions preserve the \(M_t^2\) scaling, in general.
    
    Further generalizations of NI theory treat the case of an inhomogeneous background field and are perhaps most relevant to the solar wind; these theories predict a linear scaling of \textit{rms} fluctuation strength with \(M_t\) \citep{BhattacharjeeEA1998ApJ,Hunana2010ApJ718,Zank2017ApJ835,adhikari2020ApJ,zank2021PhPtransport}, a result also expected from a linearized analysis of the MHD equations \cite{Cho&Lazarian2003MNRAS}. The density scaling has been shown to also depend on \(\beta\) \citep{Cho&Lazarian2003MNRAS}.
    %SD I'm not sure if we should say this. I think the beta dependence from Cho & Lazarian is not very robust. (If you mean the argument of a quadratic scaling at high beta and linear scaling at low beta.)
    % In general, it is clear that 
    % the nearly-incompressible component of the turbulence emerges 
    % with greater significance within a very particular parameter regime.
    
    The turbulent Mach number \(M_t \equiv \langle (\delta v)^2 \rangle^{1/2}/c_s\) is a key parameter that measures the compressibility of the turbulence, where \(\langle (\delta v)^2 \rangle^{1/2} = \delta v_{rms}\) is the \textit{rms} of the velocity fluctuations, \(\langle \cdot \rangle\) represents a suitable averaging operator, and \(c_s\) is the ion sound speed.
    Equivalently,
    
    \begin{equation}
        M_t = \frac{\delta v_{rms}}{V_A}\sqrt{\frac{2}{\beta_p \gamma}}
    \end{equation}
    can be written by making substitutions in favor of the Alfv\'en speed \(V_A\) with \(c_s = \sqrt{\gamma k_B T / m_p}\), \(\beta_p=8\pi n_p k_B T/B^2\), and \(V_A=B/\sqrt{4\pi n_p m_p}\), where \(\gamma\) is the polytropic index, \(k_B\) is the Boltzmann constant, \(T=T_p=T_e\) is the plasma temperature for protons and electrons, \(m_p\) is the proton mass, \(\beta_p\) is the ratio of thermal to magnetic pressure for protons, \(n_p\) is the proton density, and \(B\) is the magnetic field magnitude.
    
    The purpose of this study is to further examine the relationship between density fluctuations and \(M_t\).
    To better understand the nature of compressible MHD turbulence, many numerical simulations have been employed \citep{Cho&Lazarian2003MNRAS,KowalEA2007ApJ,YangEA2016PhRvE,YangEA2017PoF,ShodaEA2019ApJL,LPYangEA2019MNRAS,Makwana&Yan2020PhRvX,YangEA2021JFM,Gan2022,FuEA2022ApJ}.
    From a wave perspective,
    compressible MHD 
    turbulence involves the nonlinear interactions between three distinct MHD modes: Alfv\'en, fast, and slow modes.
    Perpendicular ion heating is mainly caused via dissipation of the Alfv\'en mode via cyclotron resonance, although fast modes can also contribute to perpendicular ion heating via the same mechanism.
    Slow modes dissipate via Landau resonance at the fluid scale, leading to parallel ion heating for low \(\beta_p\).
    These are several pathways that density fluctuations can be connected to heating and dissipation.
    % {Mention pathways for density fluctuations to originate}
    Density fluctuations can originate from either 
    (or both) MHD waves or nonlinear structures, 
    thus 
    connecting the properties of density fluctuations 
    to the 
    strength of compressibility through two potentially distinct frameworks. 
    
    \citet{FuEA2022ApJ} investigated the relationship between the \textit{rms} density fluctuation amplitude \(\langle \delta n_p^2 \rangle ^{1/2}/\langle n_p \rangle = \delta {n_p}_{rms}/\langle n_p \rangle\) and \(M_t\) by performing a series of compressible 3D MHD simulations.
    They found a linear scaling, such that \(\delta {n_p}_{rms}/\langle n_p \rangle = \alpha M_t\).
    They find that this coefficient of proportionality \(\alpha\) is dependent on cross helicity \(\sigma_c=\frac{\langle z^2_+\rangle -\langle z^2_-\rangle}{\langle z^2_+\rangle +\langle z^2_-\rangle}\), and proton beta \(\beta_p\)
    %, and the adiabatic index \(\gamma\) (\(\gamma=1.67\) for the entirety of this study)
    , where \(z_{\pm}= | \delta \bm{v} \pm \bm{b}_{\rm A}|\) are the Els\(\Ddot{a}\)sser variables, \(\delta \bm{v}\) is the fluctuating velocity vector of the solar wind, and \(\bm{b}_{\rm A} = \bm{b}/\sqrt{4\pi m_p n_p}\) is the fluctuating magnetic field vector (\(\bm{b}\)) in Alfv\'en units.
    Similar scaling studies of solar wind density fluctuations 
    have been carried out previously, e.g., \citep{matthaeus1990JGR,TuMarsch1994JGR,AdhikariEA2020ApJS}. 
    
    In the present study, we find that the scaling of normalized density fluctuations in the solar wind during PSP's perihelia (see Section \ref{sec:data} for data description and methods; see Figure \ref{fig:dens_beta_mach} for overview radial trends) varies nearly linearly with \(M_t\) (see Figure \ref{fig:dn_mt_5min_5min}).% when considering their relative standard errors.
    We also condition the results based on \(\beta\) and \(\sigma_c\), and compare with recent simulations showcasing behavior that is similarly observed (see Figure \ref{fig:filtered_dn_mt_5min_5min}).
    However, we also find that a larger ensemble average of the same quantities yield different scaling properties.
    These results are expressed in Section \ref{sec:focused_results}.
    % Overview radial trends of \(\beta\), \(M_t\), and density fluctuations are given in Appendix \ref{sec:overview_results}.
    In Section \ref{sec:discussion}, we review our results and discuss their 
    impact on compressibility studies.

\section{Parker Solar Probe Data} \label{sec:data}

We use publicly available data\footnote{Obtained from the \href{https://spdf.gsfc.nasa.gov/}{NASA Space Physics Data Facility}.} from
the first eight orbits of Parker Solar Probe (PSP), covering the time period between October 2018 to June 2021. 
Plasma data are from the Solar Probe Cup (SPC) on the SWEAP suite \citep{kasper2016SSR,case2020ApJS}. 
For all orbits, we discard SPC data when the ``general\_flag'' variable is on.
Level 3 SPC moment data \citep[see][]{case2020ApJS} are resampled to 1-s cadence using a linear interpolation.
These data are then cleaned using a Hampel filter  \citep{pearson2002hampel,bandyopadhyay2018filter,parashar2020ApJS} in the time domain, with a filtering interval of 120 seconds.
Outliers are identified as values beyond three times the local standard deviation larger than the filtering interval's median.
The outlier values are then replaced with the local median value.
%PSP's heliocentric position and heliolatitude at the time of these measurements are shown in Figure \ref{fig:position}. Perihelia are at \(35.6\ \rs\) for orbits 1 to 3, and at \(\sim 28\ \rs\) for orbits 4 and 5. PSP stays close to the ecliptic plane in its highly elliptical orbit \citep{fox2016SSR}.
%
%\begin{figure}
%    \centering
%    \includegraphics[width=.6\textwidth]{figures/position}
%    \caption{PSP's heliocentric position and heliolatitude during its first five orbits. Times shown span UTC 2018 Oct 01 to 2020 Aug 01.}
%    \label{fig:position}
%\end{figure}
%

To show the overall trend of density fluctuations from orbits 1 through 8, we smooth the 1-s cadence time series of \(n_p\) using a boxcar average over a moving window of 2-hr duration to obtain the mean density \(\langle n_p\rangle_{2{\rm hr}}\) (at 1-s cadence), denoted by the following \(\langle \cdot \rangle_{2{\rm hr}}\). 
The fluctuations in \(n_p\) are then computed as \(\delta n_p = n_p - \langle n_p\rangle_{2{\rm hr}}\), from which we compute the \textit{rms} density fluctuations \(\langle\delta n_p^2\rangle_{2{\rm hr}}^{1/2}\) at 1-s cadence. 
Both \(\langle n_p\rangle_{2{\rm hr}}\) and \(\langle\delta n_p^2\rangle_{2{\rm hr}}^{1/2}\) are then downsampled to 1-hr cadence.

\begin{figure}
    \centering
    \includegraphics[width=.47\textwidth]{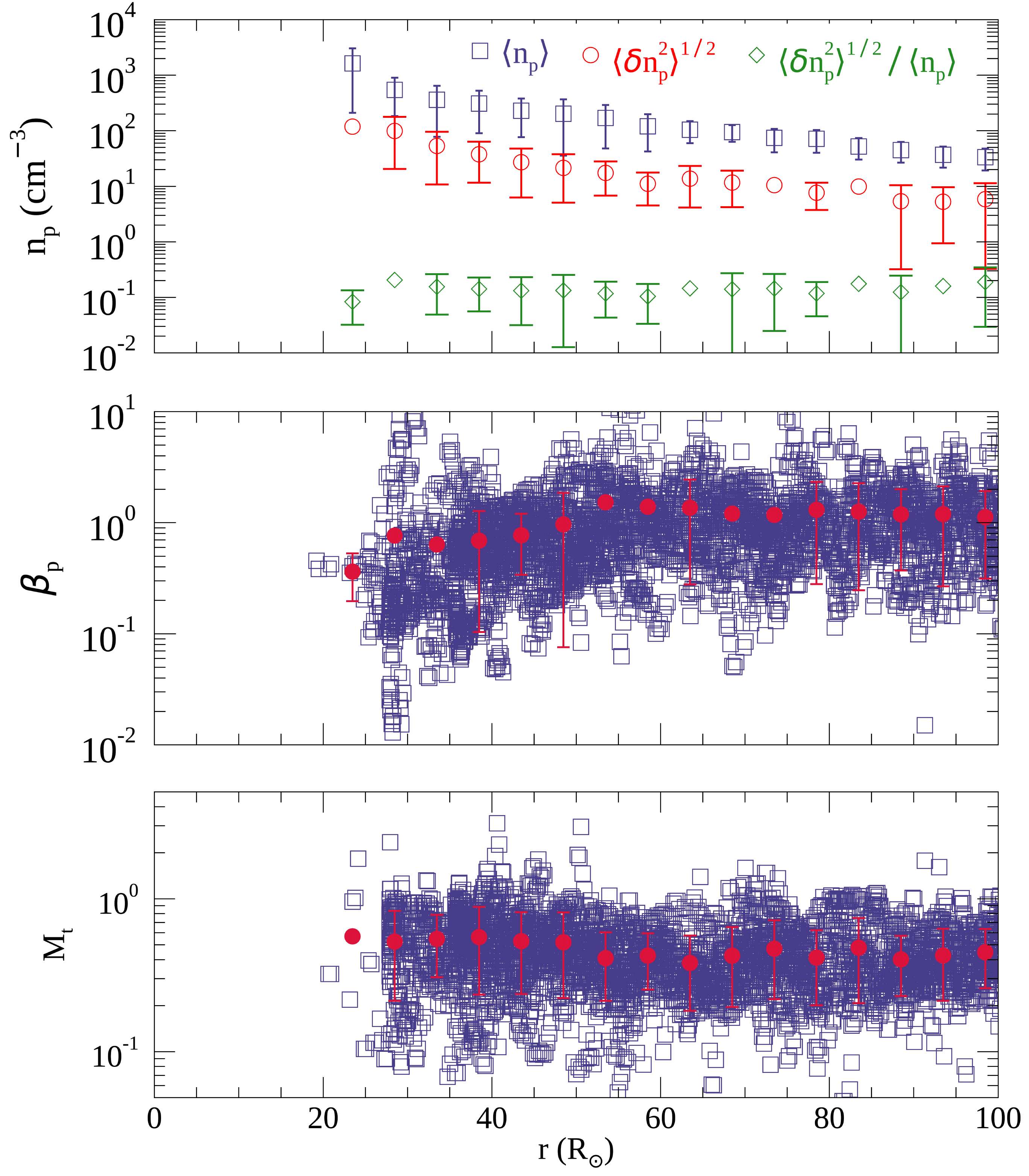}
    \caption{Overview of PSP observations in the inner heliosphere. \textit{Top}: Mean proton density \(\langle n_p\rangle\), rms density fluctuation \(\delta {n_p}_{rms}\), and the ratio \(\delta {n_p}_{rms}/\langle n_p\rangle\), plotted as a function of heliocentric distance \(r\). Data are aggregated from first eight PSP orbits and averaged within radial bins of size 5~\(\rs\).
    \textit{Middle}: Proton plasma beta \(\beta_{\text{p}}\) aggregated from first eight PSP orbits (squares), and mean values (filled red circles) within radial bins of size 5~\(\rs\). 
    \textit{Bottom}: Turbulent Mach number \(\Mt=\delta v/c_{\text{s}}\)  aggregated from first eight PSP orbits (squares), and mean  values (filled red circles) within radial bins of size 5~\(\rs\). %, where the rms velocity fluctuation is computed as \(\delta v\equiv \langle(V_R-\langle V_R\rangle)^2 + (V_T-\langle V_T\rangle)^2 + (V_N-\langle V_N\rangle)^2)\rangle^{1/2}\), and \(c_{\text{s}}\) is the sound speed, computed from the proton thermal speed \(\omega_p\) as \(c_s \equiv (5/3) \omega_p\). 
    Vertical bars represent standard deviation about corresponding means within the \(5~\rs\) bins. Note that error bars that extend to negative values are not shown on logarithmic axes. Only radial bins with at least 10 counts are included. Note that we focus on \(r \lesssim 50~\rs\)\ in our main analysis (Section \ref{sec:focused_results}).}
    \label{fig:dens_beta_mach}
\end{figure}

An overview of the density profile for all eight orbits is shown in Figure \ref{fig:dens_beta_mach}, with radially binned statistics of size 5~\(\rs\).
% 10, 5, and 2~{\(\rs\)}.
As expected, the average proton density is increasing with decreasing heliocentric distance.
The \textit{rms} density fluctuation amplitude 
is also almost steadily increasing moving inwards. 
However, the normalized  \textit{rms} density fluctuation remains rather steady, except that it 
 dips to lower values 
precipitously near 25~\(\rs\).
This might be an artifact of low sampling at these distances, requiring a larger available sampling from future PSP encounters.
Changing this radial bin size to 10 and 2~\(\rs\) produces no change in the trends.

We also provide overview trends of \(\beta_p\) and \(M_t\) in Figure \ref{fig:dens_beta_mach}, computed using a 2-hr averaging window with radial binning as described above. % of Appendix \ref{sec:overview_results}.
The turbulent Mach number ranges widely
between 0.1 and 1, with averages around 0.5 and no significant trend with radial distance.
From 100 ~\(\rs\) inward, the proton beta stays near 1 until about 60~\(\rs\), after which it decreases to an average of 0.3.
The collection of these trends point towards 
identification of a heliocentric 
region \(R<50~\rs(\sim 0.25~{\rm au})\) that may be 
meaningfully viewed as a low \(\beta_p\) and intermediate \(M_t\) environment.
Therefore, we focus on distances in this range for the results below.
% For averaging techniques used for global overview trends, refer to Appendix \ref{sec:overview_results}.

We now describe the averaging procedure used to obtain the results presented in Section \ref{sec:focused_results}, we start with the data resampled to 1-s cadence for the fundamental quantities \(n_p\), \(\bm{v}\), and \(\bm{B}\).
% , which are necessary for computing \(M_t\), \(\beta_p\), and \(\sigma_c\).
Fluctuations of these fundamental quantities are defined as \(\delta n_p = n_p - \langle n_p \rangle_1\), \(\delta \bm{v} = \bm{v} - \langle\bm{v}\rangle_1\), and \(\bm{b} = \bm{B} - \langle \bm{B} \rangle_1\), where \(\langle \cdot \rangle_1\) represents a 5-minute centered rolling average, with the averaging duration chosen to approximately match the correlation time of magnetic fluctuations observed during PSP's first several perihelia \citep[e.g.,][]{chen2020ApJS,chhiber2021ApJ_psp}.
Therefore, each fluctuating quantity is a time series with 1-s cadence.
% , which are used to compute \(M_t\), \(\beta_p\), and \(\sigma_c\), also at 1-s cadence.
We then divide these quantities into non-overlapping 5-minute sub-intervals which are used to compute a final averaged value for these quantities within each sub-interval, denoted as \(\langle \cdot \rangle_2\).
This gives the notation \(\langle \cdot \rangle = \langle \langle \cdot \rangle_1 \rangle_2\).
Finally, averages of these quantities over all non-overlapping sub-intervals
%(the full set of data points presented in Section \ref{sec:focused_results}) 
will be denoted by an overbar \(\overline{\cdot}\).
Unless otherwise specified, from here onward \(M_t\), \(\beta_p\), and \(\sigma_c\) refer to results obtained after the averaging procedure \(\langle \cdot \rangle\) is performed.
Note that \(c_s\)
can be computed from the proton thermal speed \(\omega_p\) as \(c_s \equiv \sqrt{5/3} \omega_p\), which can also be used in the framework of computing \(M_t\) and to extract a temperature (or thermal pressure).
Note that for this study, we assume that the temperature and density are equivalent for both the protons and electrons.

%--------------------------------------------------------------
\section{Results from Encounters 1-8}\label{sec:focused_results}

In this study, we examine the 
effects of \(\sigma_c\) and \(\beta_p\) 
on the relationship between \(\delta {n_p}_{rms}/\langle n_p \rangle\) and \(M_t\).
The interest here is in statistical characterization of beta, Mach number and density fluctuations of the turbulence at the energy containing scales.
(Note that the inertial range, and its properties such as the spectrum, are separate issues not investigated here.)
Therefore we average the higher resolution samples over a time value on the order of the correlation time (5-minutes) at these radial distances. 
To achieve this balance between statistical significance and regional variation, please refer to Section \ref{sec:data} for averaging techniques. 

% We smooth the 1-s resolution density data 
% using a 5-minute duration boxcar filter 
% to compute \(\delta {n_p}_{rms}\), \(\langle n_p \rangle\), \(M_t\), \(\sigma_c\), and \(\beta\) (each at 1-s cadence).
% Non-overlapping 5-minute sub-intervals are used to compute a final averaged of these quantities to compare with results from simulation.
Nine days from each encounter, centered around perihelion, are used, corresponding to a range of heliocentric distances \(0.074 < R < 0.234~{\rm au}\).
Parameter ranges for the full set of data points are as follows: \(\langle n_p \rangle \in [53,1490.5]~{\rm cm^{-3}}\) with a mean \(\overline{\langle n_p \rangle} = 334~{\rm cm^{-3}}\), \(M_t \in [0.016,1.31]\) with \(\overline{M}_t = 0.211\), \(\sigma_c \in [0.001,0.962]\) with \( \overline{\sigma}_c =0.497\), and \(\beta_p \in [0.048,18.1]\) with \(\overline{\beta}_p = 0.836\).
% Here the overbar $\overline{*}$ indicates the 
% average of the quantity $*$ over the full set of 5 minute intervals. 

\subsection{On the scaling of density fluctuations with \(M_t\)}

We investigate a possible 
power-law relationship between 
\(\delta {n_p}_{rms}/\langle n_p \rangle\) and \(M_t\).
Evaluating the standard errors in computing these two quantities \((\sigma_{\delta {n_p}_{rms}/\langle n_p \rangle} \&\ \sigma_{M_t})\), we find that \(\sigma_{\delta {n_p}_{rms}/\langle n_p \rangle} \sim \sigma_{M_t}/10\) due to the propagation of error for \(\delta {n_p}_{rms}/\langle n_p \rangle\).
We have also adjusted the non-overlapping interval duration up to 120~minutes and experienced insignificant alterations of the results presented here, other than a decrease in the number of available samples.
Based on these statistical considerations, we provide the power-law fit of \(M_t\) as a function of \(\delta {n_p}_{rms}/\langle n_p \rangle\), and invert this
relation to find an estimated 
scaling of the density fluctuations with respect to \(M_t\).
The reason for this lies within the power-law fitting method, which assumes that any error is introduced by the values of the function and not its dependent variable(s).

\begin{figure}[htp]
    \centering
    % \includegraphics[width=.45\textwidth]{figures/mecfigs/Compiled_dnbymt_5min_5min_022022.pdf}
    % \gridline{
    % \includegraphics[width=.45\textwidth]{Compiled_dnbymt_5min_5min_092022.pdf}
    \includegraphics[width=.42\textwidth]{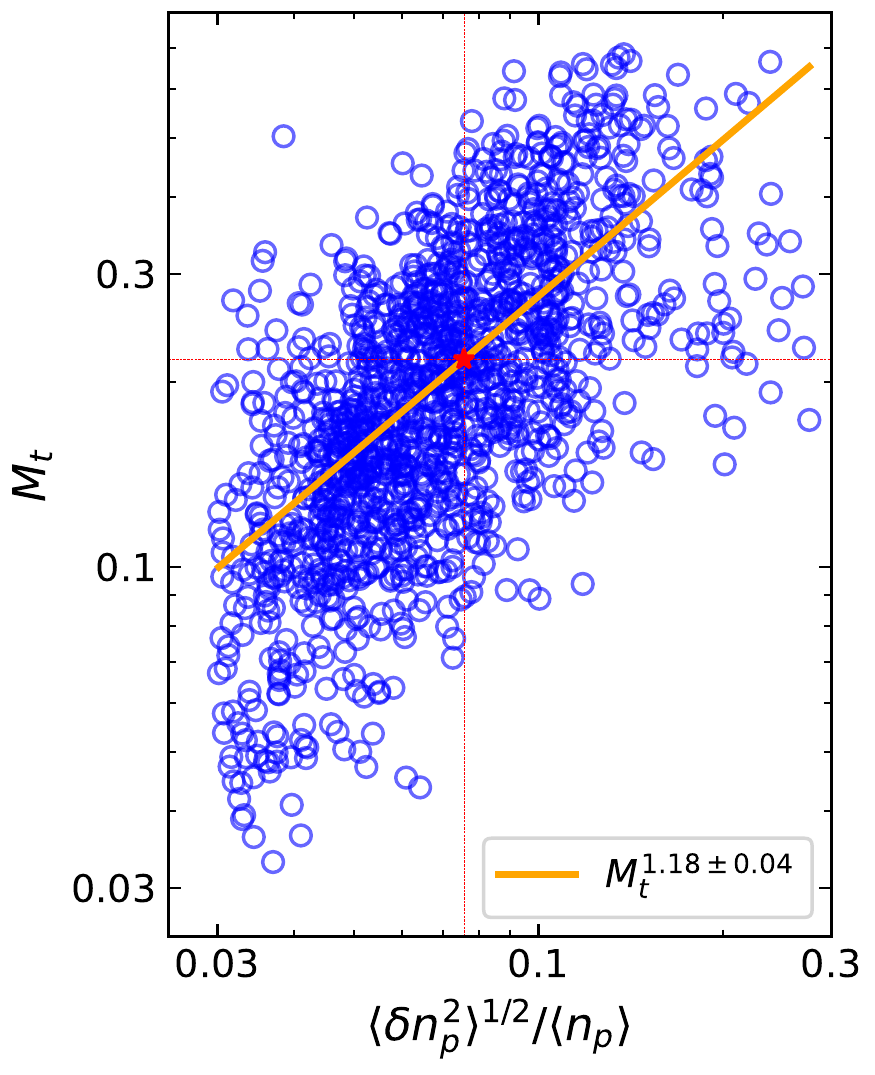}
    % }
    \caption{
    Compiled averages of \(\delta {n_p}_{rms}/\langle n_p \rangle\) and \(M_t\) spanning 9 days for encounters 1-8.  Blue circles represent 5-minute averages over non-overlapping intervals wherein statistical quantities are computed via 5 minute rolling averages (see Section \ref{sec:data} for specifics on averaging techniques).  A fitted power-law 
    % \(\delta {n_p}_{rms}/\langle n_p \rangle \sim M_t^{0.49 \pm 0.02}\) (left) or 
    is given by the solid orange line, and is 
equivalent to     \(\delta {n_p}_{rms}/\langle n_p \rangle \sim M_t^{1.18 \pm 0.04}\). The red star represents the average of \(\delta {n_p}_{rms}/\langle n_p \rangle\) and \(M_t\) over all intervals.}
    \label{fig:dn_mt_5min_5min}
\end{figure}

An overview of  \(M_t\) as a function of \(\delta {n_p}_{rms}/\langle n_p \rangle\) is shown in Figure \ref{fig:dn_mt_5min_5min}. In this locally averaged point of view, we observe that \(\delta {n_p}_{rms}/\langle n_p \rangle \sim 
%M_t^{0.49 \pm 0.02}~{\rm or}~
M_t^{1.18 \pm 0.04}\) as determined 
from the fit shown in Figure \ref{fig:dn_mt_5min_5min}. This power-law is close to the  \(M_t^{0.97}\) scaling observed by \cite{AdhikariEA2020ApJS} in PSP's first orbit. Note that we have applied constraints on our analysis, such that intervals with values of \(\delta {n_p}_{rms}/\langle n_p \rangle < 0.03\) have been discarded due to instrument noise considerations \citep{case2020ApJS}.
Additionally, 1\% of the lower and upper values of \(M_t\) have been discarded to remove possible outliers from the core results.
% Taking the average between the two fitted power-law scalings yields \(\delta {n_p}_{rms}/\langle n_p \rangle \sim M_t^{0.84 \pm 0.05}\).
    
The obtained power-law scaling is inconsistent with that predicted from NI theory for homogeneous flows (density fluctuations scale with \(M_t^2\)).
The result obtained here follows much more closely 
the elementary prediction from linear theory, or NI theory with an inhomogeneous background field (density fluctuations scale with \(M_t\)). However, significant 
statistical variations 
are seen. 
%consideration is required for this statement to be true.

Shifting to a global perspective 
for the relationship of these quantities, 
we perform averaging
over the full set of non-overlapping sub-intervals (all blue circles in Figure \ref{fig:dn_mt_5min_5min}).
We find that \(\overline{ \delta {n_p}_{rms}/\langle n_p \rangle}/\overline{M}_t~\approx~0.36\) and \(\overline{\delta {n_p}_{rms}/\langle n_p \rangle}/\overline{M}_t^2~\approx~ 1.70\).
Values of \(\overline{\delta {n_p}_{rms}/\langle n_p \rangle}\) and \(\overline{M}_t\) are shown in Figure \ref{fig:dn_mt_5min_5min} represented by the red star.
While no formal conclusion can be formed
based on these constants of proportionality, 
their values might
be seen as slightly favoring the 
homogeneous  NI theory. 
Perhaps most likely 
is that we are observing some 
mixture of inhomogeneous NI $M_t$ 
scaling and homogeneous NI $M_t^2$ scaling.

\begin{figure*}[ht!]
    \centering
    \gridline{
    % \includegraphics[width=.5\textwidth]{figures/mecfigs/Compiled_dnbymt_ch_beta_5min_5min_022022.pdf}
    % \includegraphics[scale=.5]{figures/mecfigs/Fu_Poster_Dens_Mt_ch_beta_left.png}
    % }
    \includegraphics[width=.9\textwidth]{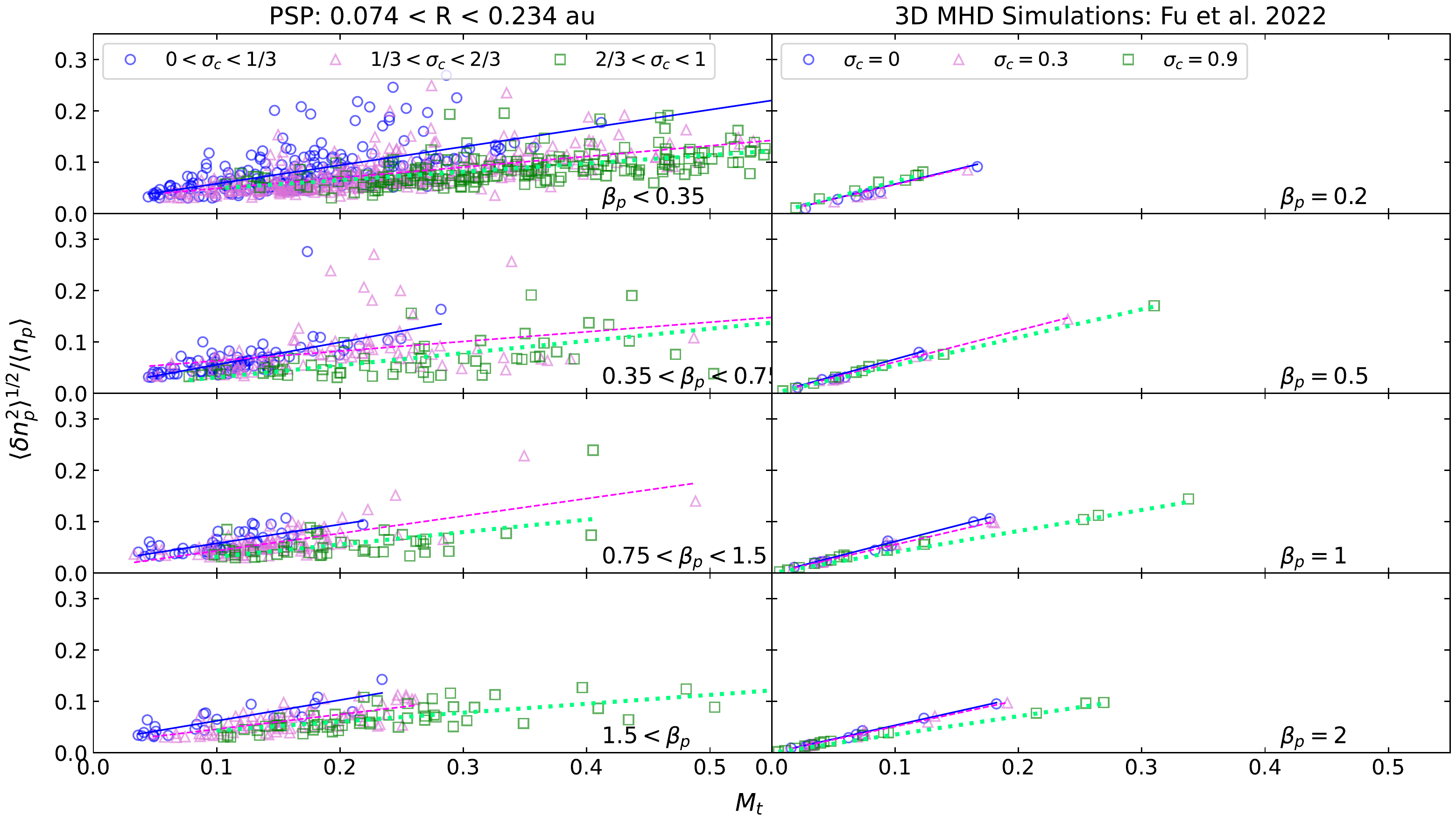}
    }
    \caption{
    (Left column) Compiled averages of density fluctuations \(\delta {n_p}_{rms}/\langle n_p \rangle\) as a function of turbulent Mach number \(M_t\) spanning 9 days each for PSP encounters 1-8.  Points represent 5-minute averages over non-overlapping intervals whose quantities are computed via 5 minute rolling averages (see Section \ref{sec:data} for specifics on averaging techniques). (Right column) \(\delta {n_p}_{rms}/\langle n_p \rangle\) as a function of \(M_t\) for individual simulations, adapted from \citet{FuEA2022ApJ}.  Both the left and right plots are conditioned on 
    ranges of cross helicity \(\sigma_c\) and proton plasma beta \(\beta_p\). Solid blue, dashed pink, and dotted green lines represent fits to blue circles, pink triangles, and green squares, respectively.
    }
    \label{fig:filtered_dn_mt_5min_5min}
\end{figure*}

\begin{table*}[ht!]
    \centering
    \begin{tabular}{c|c|c|c|c}
        \(\langle \delta n_p^2 \rangle^{1/2}/\langle n_p \rangle \sim \alpha M_t\) & \(\beta_p < 0.35\) & \(0.35 < \beta_p < 0.75\) & \(0.75 < \beta_p < 1.50\) & \(1.50 < \beta_p\) \\
        \hline
        \(0 < | \sigma_c | < 1/3\)   & 0.36 $\pm$ 0.03 & 0.44 $\pm$ 0.06 & 0.37 $\pm$ 0.05 & 0.41 $\pm$ 0.05 \\
        \(1/3 < | \sigma_c | < 2/3\) & 0.21 $\pm$ 0.01 & 0.19 $\pm$ 0.05 & 0.34 $\pm$ 0.03 & 0.29 $\pm$ 0.03 \\
        \(2/3 < | \sigma_c | < 1\)   & 0.16 $\pm$ 0.01 & 0.24 $\pm$ 0.04 & 0.24 $\pm$ 0.05 & 0.17 $\pm$ 0.02 \\
        % \(0 < | \sigma_c | < 1/3\)   & 0.36 $\pm$ 0.032 & 0.44 $\pm$ 0.058 & 0.37 $\pm$ 0.054 & 0.41 $\pm$ 0.050 \\
        % \(1/3 < | \sigma_c | < 2/3\) & 0.21 $\pm$ 0.013 & 0.19 $\pm$ 0.045 & 0.34 $\pm$ 0.033 & 0.29 $\pm$ 0.033 \\
        % \(2/3 < | \sigma_c | < 1\)   & 0.16 $\pm$ 0.014 & 0.24 $\pm$ 0.037 & 0.24 $\pm$ 0.054 & 0.17 $\pm$ 0.023 \\
        %  &  &  &  &  \\
    \end{tabular}
    \caption{Linear fits and standard errors (\(\alpha \pm \sigma_{\alpha}\)) corresponding to the lines in Figure \ref{fig:filtered_dn_mt_5min_5min}, which are conditioned on cross helicity \(\sigma_c\) (rows) and plasma beta \(\beta\) (columns). 
    % {\color{red} CONSTRAINTS: Mt with 1\% of outliers removed AND dn/n>.03.}
    }
    \label{tab:lin_fits_5min_5min}
\end{table*}

\subsection{Influence of plasma beta and cross helicity on the scaling}

We also have examined the linear dependence of 
density fluctuations when separately grouping intervals conditioned on values of the parameters \(\sigma_c\) and \(\beta_p\). 
Specifically we examine how these properties affect the scaling of \(\delta {n_p}_{rms}/\langle n_p \rangle\) with \(M_t\).
There may be some \(\gamma\) dependence, which is difficult to define from the data \citep[however, see][]{TottenEA1995JGR,Livadiotis2018Entropy,Nicolaou&Libadiotis2019Entropy,NicolaouEA2020ApJ}.
Therefore, we set \(\gamma=1.67\) for the entirety of this study, noting that any change to \(\gamma\) will not change any qualitative remarks on the observed conditioned trends.
We compare our results with similar analyses based on compressible 3D MHD turbulence simulations performed by \citep{FuEA2022ApJ}.
% We also can compare these conditioned observational results with results from 
% a number of numerical simulations that have similar parameters. 
This will 
provide potential confirmation of observed effects when \(\beta_p\) and \(\sigma_c\) are varied. 
These conditioned in-situ (left panel) and simulation (right panel) results are given in Figure \ref{fig:filtered_dn_mt_5min_5min}.
Since the overall results are well-fit by a linear relationship between density fluctuations and Mach number (Figure \ref{fig:dn_mt_5min_5min}), we perform linear fits \(\delta {n_p}_{rms}/\langle n_p \rangle = \alpha M_t\), represented by the straight lines in the left panel of Figure \ref{fig:filtered_dn_mt_5min_5min}, with details of the fits provided in Table \ref{tab:lin_fits_5min_5min}.
These conditioned results provide observations of different systematic changes for different combined ranges of \(\sigma_c\) and \(\beta_p\).

Specific conclusions are:
\begin{itemize}
    \item For fixed \(\sigma_c\), we observe no significant trend of linear coefficient \(\alpha\) with respect to \(\beta_p\).
    %\item For intermediate to high \(\sigma_c\), we observe no significant trend w.r.t. \(\beta_p\).
    \item For fixed \(\beta_p\), we observe a decreasing \(\alpha\) with increasing \(\sigma_c\).
\end{itemize}
The description of these trends can be confirmed numerically via the linear least-squares fit to the function \(\delta {n_p}_{rms}/\langle n_p \rangle \sim \alpha M_t\) in Table \ref{tab:lin_fits_5min_5min} and comparing the value of \(\alpha\) over different pairings of \(\sigma_c\) and \(\beta_p\).
Further discussion of these trends can be found in Section \ref{sec:discussion}.
%Note that in this approach 
%a linear scaling is assumed to be appropriate.

Next, we compare these trends with results based on 3D MHD simulations of compressible turbulence published by \cite{FuEA2022ApJ}. 
These results utilized the high-performance code ATHENA++ \citep{StoneEA2020ApJS} to solve the ideal compressible MHD equations within an elongated box domain with periodic boundary conditions.
The system started from a uniform mass density and uniform background magnetic field with zero fluctuations.
Plasma parameters are selected to represent typical solar wind conditions found within 1~au. 
Further details on the simulations may be found in \cite{FuEA2022ApJ}.

The aforementioned observed trends are also seen in simulations  \citep{FuEA2022ApJ}, in which several runs of different \(M_t\) and \(M_A\) are used to compute a density fluctuation amplitude over the simulation domain for fixed \(\sigma_c\) and \(\beta_p\). 
The results with $\gamma=1.67$ from \citet{FuEA2022ApJ} are shown on the right panel of Figure \ref{fig:filtered_dn_mt_5min_5min}.
Here we note two caveats when comparing PSP observations to these simulations.
The first is that the PSP observations correspond to averaging at around the energy injection scale, where as simulation results presented in Figure \ref{fig:filtered_dn_mt_5min_5min} are computed for scales in the inertial range, although including the energy injection scales do not affect these trends.
% \textbf{Secondly, PSP sampling of the solar wind is strongly biased parallel to the background magnetic field.
% However, results from the 3D MHD simulations are produced by averaging over all available directions (perpendicular and parallel to the background magnetic field).
% Although a better quantitative comparison can be made with taking proper directional averaging over the simulation domain \citep{DuEA2023ApJ}, qualitative trends should remain consistent.}
%SD My suggestion as follows.
Secondly, results from the 3D MHD simulations are produced by global averaging over a 3D volume. Such averaging is not equivalent to spacecraft sampling along a 1D trajectory, which contains more random variations and may have a dependence on the angle between the sample trajectory and background magnetic field due to anisotropy of fluctuations \citep{DuEA2023ApJ}. A systematic bias can be introduced since PSP favors sampling of the solar wind parallel to the background magnetic field. However, since sampling likely affects the calculation of both $M_t$ and $\delta {n_p}_{rms}/ \langle n_p\rangle$ similarly, we expect the qualitative trends in the scaling relation to remain consistent even if proper directional averaging is made.

% \begin{table*}[ht!]
%     \centering
%     \begin{tabular}{c|c|c|c|c}
%         \(\langle \delta n_p^2 \rangle^{1/2}/\langle n_p \rangle \sim M_t^{\kappa}\) & \(\beta < 0.35\) & \(0.35 < \beta < 0.75\) & \(0.75 < \beta < 1.50\) & \(1.50 < \beta\) \\
%         \hline
%         \(0 < | \sigma_c | < 1/3\)   & 0.67 $\pm$ 0.053 & 0.68 $\pm$ 0.070 & 0.55 $\pm$ 0.086 & 0.62 $\pm$ 0.079 \\
%         \(1/3 < | \sigma_c | < 2/3\) & 0.61 $\pm$ 0.033 & 0.52 $\pm$ 0.078 & 0.61 $\pm$ 0.070 & 0.67 $\pm$ 0.073 \\
%         \(2/3 < | \sigma_c | < 1\)   & 0.60 $\pm$ 0.049 & 0.63 $\pm$ 0.111 & 0.57 $\pm$ 0.134 & 0.67 $\pm$ 0.080 \\
%         %  &  &  &  &  \\
%     \end{tabular}
%     \caption{Powerlaw fits and standard errors (\(\kappa \pm \sigma_{\kappa}\)) corresponding to the curves in Figure \ref{fig:filtered_dn_mt_5min_5min}, which are filtered by cross helicity \(\sigma_c\) (rows) and plasma beta \(\beta\) (columns). {\color{red} CONSTRAINTS: Mt with 1\% of outliers removed AND dn/n>.03.}}
%     \label{tab:powerlaw_fits_5min_5min}
% \end{table*}

\section{Discussion}\label{sec:discussion}

We have examined the relationship between the normalized proton density variation and 
the turbulent Mach number, a problem that has been examined previously using several different spacecraft 
\citep{matthaeus1990JGR,TuMarsch1994JGR} in different regions of the heliosphere, with \citet{AdhikariEA2020ApJS} utilizing PSP's first encounter. 
Here we extend these studies to PSP's dataset of the first eight encounters.   
%We find that the observed scaling behavior of density fluctuations with turbulent Mach number in the inner heliosphere near PSP's perihelia 
%depends significantly on the associated statistical variability in each parameter.
% \footnote{Statistical errors as well as measurement errors can be considered. The statistical errors are believed to be dominant in most of the data.}
%In response to the standard error in the turbulent Mach number being larger that the standard error in the normalized \textit{rms} density fluctuations, 
%We compute  a power-law fit to \(M_t\) as a function of \(\delta {n_p}_{rms}/\langle n_p \rangle\) (see Figure \ref{fig:dn_mt_5min_5min}).
%We find that
% \(\delta {n_p}_{rms} \sim M_t^{0.49 \pm 0.02}\) and 
%\(M_t \sim \left( \delta {n_p}_{rms} /\langle n_p \rangle \right)^{0.85 \pm 0.03}\), which is equivalent to 
We find a power-law scaling of \(\delta {n_p}_{rms} /\langle n_p \rangle \sim M_t^{1.18 \pm 0.04}\), consistent  with NI theory extended to inhomogeneous background fields \citep{BhattacharjeeEA1998ApJ,Hunana2010ApJ718}, and also with the prediction based on linearized MHD equations. Since the best fit lies between the compressible wave scaling $\sim M_t$ and the NI scaling $\sim M_t^2$ expected for a homogeneous background field \citep{zank1993nearly},
it is also possible to interpret the observed result as representing a mixture of these two types of plasma states.
%On the other hand, the \(\delta {n_p}_{rms}/\langle n_p \rangle \sim M_t^2\) scaling expected for NI scaling of homogeneous flows \citep{matthaeus1988PoF,MatthaeusEA1991JGR,zank1993nearly}, is a somewhat less good fit. 
We also examined the influence of 
the averaging domain on the relationship between 
density fluctuations and turbulent Mach number. When 
averaging over the full set of PSP intervals, we find that the coefficient of the \(\sim M_t^2\) scaling is closer to unity than that of the $\sim M_t$ scaling, 
suggesting some relevance of the homogeneous 
NI theory.
% However many of the observed PSP results would also lie quite close to a $M_t^2$ trend line. 

%The intervals investigated are largely dominated by \(\beta_p < 0.7\) and \(\sigma_c > 0.33\).

% The known effects of plasma beta and cross helicity on normalized density fluctuations, as deduced from simulations, are recovered in observations when data is conditioned into subsets relative to \(\sigma_c\) and \(\beta_p\).

Another interesting point of discussion is the relationship of the present results to both 
incompressive cascade rate laws \citep{politano1998GRL} and compressive cascade rate laws \citep{hadid2017energy}.
The empirical scaling found in the present paper may provide guidance as to which of these formulations may be most appropriate for application to solar wind. Moreover, analyses that employ the compressive cascade rate formalism, such as \citet{hadid2017energy} 
infer an empirical relationship between cascade rate and turbulent Mach number, i.e., $\epsilon \sim M^{2.67}$ for slow wind and $\epsilon \sim M^{1.5}$ for fast wind. Using our result that $\delta \rho \sim M^{1.18}$, and assuming the results can be combined, one may 
deduce a relationship between $\delta \rho$ and $\epsilon$ that may be interesting for future investigation. The same observational treatment 
finds a scaling between Mach number and internal energy in an {\it isothermal} approximation, an approximation that may not be optimal for the solar wind. 
Future studies might 
employ the present empirical results to delve more deeply into various forms of NI theory 
and cascade laws to provide additional constraints 
on the underlying theory. 
Such considerations are well beyond the present intended scope.

We also carry out a procedure 
similar 
to \citet{FuEA2022ApJ} to examine the effects of plasma beta  and cross helicity on the linear dependence of density fluctuations on \(M_t\) and compare our observational analysis with their results from 3D MHD simulations (see Figure \ref{fig:filtered_dn_mt_5min_5min}).
From the conditioned 
results using PSP observations 
and in comparison to their simulation results, we find a consistency in the trends with increasing \(\sigma_c\) for a given range of \(\beta_p\) (see Table \ref{tab:lin_fits_5min_5min}).
One clear result is that when cross helicity is increased, 
the normalized density fluctuations are decreased.
This is generally consistent with the idea that Alfv\'enic fluctuations are incompressive.
It is rather significant that PSP observations in the compressible solar wind reflects the same systematic changes observed in compressible three-dimensional MHD simulations.

One might want to develop a special treatment of density fluctuations for regions of solar wind having low plasma beta ($\ll 1$), a condition that occurs sometimes, especially in magnetic clouds \citep{smith2006interplanetary} but is not common in general. Low plasma beta has been recently treated using weak turbulence theory based on fast magnetosonic modes \citep{galtier2023fast}. However the absence of Alfv\'en modes in that treatment makes it unlikely to explain the observations in the present paper.  Given that Alfv\'enic fluctuations are typically present \citep{chen2020ApJS}, a relevant theory related to the generation of density fluctuations is the parametric decay instability \citep[e.g.,][]{FuEA2018ApJ} of Alfv\'en waves. However this mechanism becomes more important at low plasma beta when the fluctuation is coherent, which, again, is not typical for the observations we have presented.

Anisotropy in the inner heliosphere due to a more dominantly radial magnetic field has been found to affect many properties of turbulence, such as differences in 
magnetic power spectra, correlation lengths, and heating when decomposing these quantities into parallel and perpendicular components.
A possible influence on the observed distributions
and associated 
scalings
is that these intervals are largely dominated by parallel sampling by PSP \citep{CuestaEA2022ApJLIsotropization}.
Anisotropy can cause density fluctuations sampled along the mean magnetic field to be weaker compared to the fluctuations sampled transverse to the mean field \citep{DuEA2023ApJ}.
Further analysis along the lines of 
prior investigations
(e.g., 
\citet{matthaeus1990JGR,ZankEA1990GeoRLDensity,matthaeus1996jgr,DassoEA2005ApJL}) would be 
required to examine further the relationship between 
the well-studied types of anisotropy and the corresponding observed 
properties of the density fluctuations
\citep{ChhiberEA2023inprep}. 
%Rohit, Victoria et al, in prep.
Another point of interest is to investigate these results in regards to solar wind speed.
Although future PSP orbits may encounter more frequent faster solar wind speeds, the intervals in this study favor slow wind speeds (\(<450~{\rm km/s}\)) by nearly 98\%.

\section{Acknowledgements}

    This research is partially supported by NASA under the Heliophysics Supporting Research program grants
    80NSSC18K1210 and 80NSSC18K1648, by the Parker Solar Probe Guest Investigator program 80NSSC21K1765 at the University of Delaware, 
    by Heliophysics Guest Investigator program 80NSSC19K0284,
    and the PUNCH project under subcontract 
    NASA/SWRI N99054DS.
    In collaboration with Los Alamos National Laboratory (LANL), and partially supported by a NASA LWS grant 80NSSC20K0377(subcontract 655-001), those affiliated with University of Delaware visited LANL 
    in summer 2022 to work on this project. 
    S. Du and H. Li acknowledge the support by DOE OFES program and LANL/LDRD program. Z. Gan and X. Fu are supported by NASA under Award No. 80NSSC20K0377.

%-------------------------------------------------------------
% \bibliography{chhibref}

\begin{thebibliography}{}
\expandafter\ifx\csname natexlab\endcsname\relax\def\natexlab#1{#1}\fi
\providecommand{\url}[1]{\href{#1}{#1}}
\providecommand{\dodoi}[1]{doi:~\href{http://doi.org/#1}{\nolinkurl{#1}}}
\providecommand{\doeprint}[1]{\href{http://ascl.net/#1}{\nolinkurl{http://ascl.net/#1}}}
\providecommand{\doarXiv}[1]{\href{https://arxiv.org/abs/#1}{\nolinkurl{https://arxiv.org/abs/#1}}}

\bibitem[{{Adhikari} {et~al.}(2020{\natexlab{a}}){Adhikari}, {Zank}, \&
  {Zhao}}]{adhikari2020ApJ}
{Adhikari}, L., {Zank}, G.~P., \& {Zhao}, L.~L. 2020{\natexlab{a}}, \apj, 901,
  102, \dodoi{10.3847/1538-4357/abb132}

\bibitem[{{Adhikari} {et~al.}(2020{\natexlab{b}}){Adhikari}, {Zank}, {Zhao},
  {Kasper}, {Korreck}, {Stevens}, {Case}, {Whittlesey}, {Larson}, {Livi}, \&
  {Klein}}]{AdhikariEA2020ApJS}
{Adhikari}, L., {Zank}, G.~P., {Zhao}, L.~L., {et~al.} 2020{\natexlab{b}},
  \apjs, 246, 38, \dodoi{10.3847/1538-4365/ab5852}

\bibitem[{{Bandyopadhyay} {et~al.}(2018){Bandyopadhyay}, {Chasapis}, {Chhiber},
  {Parashar}, {Maruca}, {Matthaeus}, {Schwartz}, {Eriksson}, {Le Contel},
  {Breuillard}, {Burch}, {Moore}, {Pollock}, {Giles}, {Paterson}, {Dorelli},
  {Gershman}, {Torbert}, {Russell}, \& {Strangeway}}]{bandyopadhyay2018filter}
{Bandyopadhyay}, R., {Chasapis}, A., {Chhiber}, R., {et~al.} 2018, \apj, 866,
  81, \dodoi{10.3847/1538-4357/aade93}

\bibitem[{{Bavassano} \& {Bruno}(1995)}]{BavassanoBruno1995JGRDensity}
{Bavassano}, B., \& {Bruno}, R. 1995, \jgr, 100, 9475,
  \dodoi{10.1029/94JA03048}

\bibitem[{Bayly {et~al.}(1992)Bayly, Levermore, \& Passot}]{BaylyEA92}
Bayly, B.~J., Levermore, C.~D., \& Passot, T. 1992, Physics of Fluids A: Fluid
  Dynamics, 4, 945, \dodoi{10.1063/1.858275}

\bibitem[{{Bhattacharjee} {et~al.}(1998){Bhattacharjee}, {Ng}, \&
  {Spangler}}]{BhattacharjeeEA1998ApJ}
{Bhattacharjee}, A., {Ng}, C.~S., \& {Spangler}, S.~R. 1998, \apj, 494, 409,
  \dodoi{10.1086/305184}

\bibitem[{{Case} {et~al.}(2020){Case}, {Kasper}, {Stevens}, {Korreck},
  {Paulson}, {Daigneau}, {Caldwell}, {Freeman}, {Henry}, {Klingensmith},
  {Bookbinder}, {Robinson}, {Berg}, {Tiu}, {Wright}, {Reinhart}, {Curtis},
  {Ludlam}, {Larson}, {Whittlesey}, {Livi}, {Klein}, \&
  {Martinovi{\'c}}}]{case2020ApJS}
{Case}, A.~W., {Kasper}, J.~C., {Stevens}, M.~L., {et~al.} 2020, \apjs, 246,
  43, \dodoi{10.3847/1538-4365/ab5a7b}

\bibitem[{{Chen} {et~al.}(2020){Chen}, {Bale}, {Bonnell}, {Borovikov}, {Bowen},
  {Burgess}, {Case}, {Chandran}, {de Wit}, {Goetz}, {Harvey}, {Kasper},
  {Klein}, {Korreck}, {Larson}, {Livi}, {MacDowall}, {Malaspina}, {Mallet},
  {McManus}, {Moncuquet}, {Pulupa}, {Stevens}, \& {Whittlesey}}]{chen2020ApJS}
{Chen}, C.~H.~K., {Bale}, S.~D., {Bonnell}, J.~W., {et~al.} 2020, \apjs, 246,
  53, \dodoi{10.3847/1538-4365/ab60a3}

\bibitem[{{Chhiber} {et~al.}(2021){Chhiber}, {Usmanov}, {Matthaeus}, \&
  {Goldstein}}]{chhiber2021ApJ_psp}
{Chhiber}, R., {Usmanov}, A.~V., {Matthaeus}, W.~H., \& {Goldstein}, M.~L.
  2021, \apj, 923, 89, \dodoi{10.3847/1538-4357/ac1ac7}

\bibitem[{{Chhiber} {et~al.}(2019){Chhiber}, {Usmanov}, {Matthaeus},
  {Parashar}, \& {Goldstein}}]{chhiber2019psp2}
{Chhiber}, R., {Usmanov}, A.~V., {Matthaeus}, W.~H., {Parashar}, T.~N., \&
  {Goldstein}, M.~L. 2019, \apjs, 242, 12, \dodoi{10.3847/1538-4365/ab16d7}

\bibitem[{{Cho} \& {Lazarian}(2003)}]{Cho&Lazarian2003MNRAS}
{Cho}, J., \& {Lazarian}, A. 2003, \mnras, 345, 325,
  \dodoi{10.1046/j.1365-8711.2003.06941.x}

\bibitem[{{Cuesta} {et~al.}(2022){Cuesta}, {Chhiber}, {Roy}, {Goodwill},
  {Pecora}, {Jarosik}, {Matthaeus}, {Parashar}, \&
  {Bandyopadhyay}}]{CuestaEA2022ApJLIsotropization}
{Cuesta}, M.~E., {Chhiber}, R., {Roy}, S., {et~al.} 2022, \apjl, 932, L11,
  \dodoi{10.3847/2041-8213/ac73fd}

\bibitem[{{Dasso} {et~al.}(2005){Dasso}, {Milano}, {Matthaeus}, \&
  {Smith}}]{DassoEA2005ApJL}
{Dasso}, S., {Milano}, L.~J., {Matthaeus}, W.~H., \& {Smith}, C.~W. 2005,
  \apjl, 635, L181, \dodoi{10.1086/499559}

\bibitem[{{Du} {et~al.}(2023){Du}, {Li}, {Gan}, \& {Fu}}]{DuEA2023ApJ}
{Du}, S., {Li}, H., {Gan}, Z., \& {Fu}, X. 2023, \apj, 946, 74,
  \dodoi{10.3847/1538-4357/acc10b}

\bibitem[{{Fu} {et~al.}(2022){Fu}, {Li}, {Gan}, {Du}, \&
  {Steinberg}}]{FuEA2022ApJ}
{Fu}, X., {Li}, H., {Gan}, Z., {Du}, S., \& {Steinberg}, J. 2022, \apj, 936,
  127, \dodoi{10.3847/1538-4357/ac8802}

\bibitem[{{Fu} {et~al.}(2018){Fu}, {Li}, {Guo}, {Li}, \&
  {Roytershteyn}}]{FuEA2018ApJ}
{Fu}, X., {Li}, H., {Guo}, F., {Li}, X., \& {Roytershteyn}, V. 2018, \apj, 855,
  139, \dodoi{10.3847/1538-4357/aaacd6}

\bibitem[{Galtier(2023)}]{galtier2023fast}
Galtier, S. 2023, Journal of Plasma Physics, 89, 905890205

\bibitem[{{Gan} {et~al.}(2022){Gan}, {Li}, {Fu}, \& {Du}}]{Gan2022}
{Gan}, Z., {Li}, H., {Fu}, X., \& {Du}, S. 2022, \apj, 926, 222,
  \dodoi{10.3847/1538-4357/ac4d9d}

\bibitem[{{Grappin} {et~al.}(1990){Grappin}, {Mangeney}, \&
  {Marsch}}]{GrappinEA1990}
{Grappin}, R., {Mangeney}, A., \& {Marsch}, E. 1990, \jgr, 95, 8197,
  \dodoi{10.1029/JA095iA06p08197}

\bibitem[{Hadid {et~al.}(2017)Hadid, Sahraoui, \& Galtier}]{hadid2017energy}
Hadid, L., Sahraoui, F., \& Galtier, S. 2017, The Astrophysical Journal, 838, 9

\bibitem[{{Hunana} \& {Zank}(2010)}]{Hunana2010ApJ718}
{Hunana}, P., \& {Zank}, G.~P. 2010, \apj, 718, 148,
  \dodoi{10.1088/0004-637X/718/1/148}

\bibitem[{{Kasper} {et~al.}(2016){Kasper}, {Abiad}, {Austin}, {Balat-Pichelin},
  {Bale}, {Belcher}, {Berg}, {Bergner}, {Berthomier}, {Bookbinder}, {Brodu},
  {Caldwell}, {Case}, {Chandran}, {Cheimets}, {Cirtain}, {Cranmer}, {Curtis},
  {Daigneau}, {Dalton}, {Dasgupta}, {DeTomaso}, {Diaz-Aguado}, {Djordjevic},
  {Donaskowski}, {Effinger}, {Florinski}, {Fox}, {Freeman}, {Gallagher},
  {Gary}, {Gauron}, {Gates}, {Goldstein}, {Golub}, {Gordon}, {Gurnee}, {Guth},
  {Halekas}, {Hatch}, {Heerikuisen}, {Ho}, {Hu}, {Johnson}, {Jordan},
  {Korreck}, {Larson}, {Lazarus}, {Li}, {Livi}, {Ludlam}, {Maksimovic},
  {McFadden}, {Marchant}, {Maruca}, {McComas}, {Messina}, {Mercer}, {Park},
  {Peddie}, {Pogorelov}, {Reinhart}, {Richardson}, {Robinson}, {Rosen},
  {Skoug}, {Slagle}, {Steinberg}, {Stevens}, {Szabo}, {Taylor}, {Tiu}, {Turin},
  {Velli}, {Webb}, {Whittlesey}, {Wright}, {Wu}, \& {Zank}}]{kasper2016SSR}
{Kasper}, J.~C., {Abiad}, R., {Austin}, G., {et~al.} 2016, \ssr, 204, 131,
  \dodoi{10.1007/s11214-015-0206-3}

\bibitem[{{Kasper} {et~al.}(2021){Kasper}, {Klein}, {Lichko}, {Huang}, {Chen},
  {Badman}, {Bonnell}, {Whittlesey}, {Livi}, {Larson}, {Pulupa}, {Rahmati},
  {Stansby}, {Korreck}, {Stevens}, {Case}, {Bale}, {Maksimovic}, {Moncuquet},
  {Goetz}, {Halekas}, {Malaspina}, {Raouafi}, {Szabo}, {MacDowall}, {Velli},
  {Dudok de Wit}, \& {Zank}}]{KasperEAprl2021}
{Kasper}, J.~C., {Klein}, K.~G., {Lichko}, E., {et~al.} 2021, \prl, 127,
  255101, \dodoi{10.1103/PhysRevLett.127.255101}

\bibitem[{{Klein} {et~al.}(1993){Klein}, {Bruno}, {Bavassano}, \&
  {Rosenbauer}}]{KleinEA1993JGRDensity}
{Klein}, L., {Bruno}, R., {Bavassano}, B., \& {Rosenbauer}, H. 1993, \jgr, 98,
  7837, \dodoi{10.1029/92JA02906}

\bibitem[{{Kowal} {et~al.}(2007){Kowal}, {Lazarian}, \&
  {Beresnyak}}]{KowalEA2007ApJ}
{Kowal}, G., {Lazarian}, A., \& {Beresnyak}, A. 2007, \apj, 658, 423,
  \dodoi{10.1086/511515}

\bibitem[{{Livadiotis}(2018)}]{Livadiotis2018Entropy}
{Livadiotis}, G. 2018, Entropy, 20, 799, \dodoi{10.3390/e20100799}

\bibitem[{{Makwana} \& {Yan}(2020)}]{Makwana&Yan2020PhRvX}
{Makwana}, K.~D., \& {Yan}, H. 2020, Physical Review X, 10, 031021,
  \dodoi{10.1103/PhysRevX.10.031021}

\bibitem[{{Malara} {et~al.}(1996){Malara}, {Primavera}, \&
  {Veltri}}]{MalaraEAJGR1996}
{Malara}, F., {Primavera}, L., \& {Veltri}, P. 1996, \jgr, 101, 21597,
  \dodoi{10.1029/96JA01637}

\bibitem[{{Matthaeus} \& {Brown}(1988)}]{matthaeus1988PoF}
{Matthaeus}, W.~H., \& {Brown}, M.~R. 1988, Physics of Fluids, 31, 3634,
  \dodoi{10.1063/1.866880}

\bibitem[{{Matthaeus} {et~al.}(1996){Matthaeus}, {Ghosh}, {Oughton}, \&
  {Roberts}}]{matthaeus1996jgr}
{Matthaeus}, W.~H., {Ghosh}, S., {Oughton}, S., \& {Roberts}, D.~A. 1996, \jgr,
  101, 7619, \dodoi{10.1029/95JA03830}

\bibitem[{{Matthaeus} {et~al.}(1990){Matthaeus}, {Goldstein}, \&
  {Roberts}}]{matthaeus1990JGR}
{Matthaeus}, W.~H., {Goldstein}, M.~L., \& {Roberts}, D.~A. 1990, \jgr, 95,
  20673, \dodoi{10.1029/JA095iA12p20673}

\bibitem[{{Matthaeus} {et~al.}(1991){Matthaeus}, {Klein}, {Ghosh}, \&
  {Brown}}]{MatthaeusEA1991JGR}
{Matthaeus}, W.~H., {Klein}, L.~W., {Ghosh}, S., \& {Brown}, M.~R. 1991, \jgr,
  96, 5421, \dodoi{10.1029/90JA02609}

\bibitem[{{Montgomery} {et~al.}(1987){Montgomery}, {Brown}, \&
  {Matthaeus}}]{MontgomeryEA1987JGR}
{Montgomery}, D., {Brown}, M.~R., \& {Matthaeus}, W.~H. 1987, \jgr, 92, 282,
  \dodoi{10.1029/JA092iA01p00282}

\bibitem[{{Nicolaou} {et~al.}(2019){Nicolaou}, {Livadiotis}, \&
  {Wicks}}]{Nicolaou&Libadiotis2019Entropy}
{Nicolaou}, G., {Livadiotis}, G., \& {Wicks}, R.~T. 2019, Entropy, 21, 997,
  \dodoi{10.3390/e21100997}

\bibitem[{{Nicolaou} {et~al.}(2020){Nicolaou}, {Livadiotis}, {Wicks},
  {Verscharen}, \& {Maruca}}]{NicolaouEA2020ApJ}
{Nicolaou}, G., {Livadiotis}, G., {Wicks}, R.~T., {Verscharen}, D., \&
  {Maruca}, B.~A. 2020, \apj, 901, 26, \dodoi{10.3847/1538-4357/abaaae}

\bibitem[{{Parashar} {et~al.}(2020){Parashar}, {Goldstein}, {Maruca},
  {Matthaeus}, {Ruffolo}, {Bandyopadhyay}, {Chhiber}, {Chasapis}, {Qudsi},
  {Vech}, {Roberts}, {Bale}, {Bonnell}, {de Wit}, {Goetz}, {Harvey},
  {MacDowall}, {Malaspina}, {Pulupa}, {Kasper}, {Korreck}, {Case}, {Stevens},
  {Whittlesey}, {Larson}, {Livi}, {Velli}, \& {Raouafi}}]{parashar2020ApJS}
{Parashar}, T.~N., {Goldstein}, M.~L., {Maruca}, B.~A., {et~al.} 2020, \apjs,
  246, 58, \dodoi{10.3847/1538-4365/ab64e6}

\bibitem[{Pearson(2002)}]{pearson2002hampel}
Pearson, R.~K. 2002, IEEE Transactions on Control Systems Technology, 10, 55,
  \dodoi{10.1109/87.974338}

\bibitem[{{Politano} \& {Pouquet}(1998)}]{politano1998GRL}
{Politano}, H., \& {Pouquet}, A. 1998, \grl, 25, 273, \dodoi{10.1029/97GL03642}

\bibitem[{{Roberts} {et~al.}(1990){Roberts}, {Goldstein}, \&
  {Klein}}]{roberts1990JGR}
{Roberts}, D.~A., {Goldstein}, M.~L., \& {Klein}, L.~W. 1990, \jgr, 95, 4203,
  \dodoi{10.1029/JA095iA04p04203}

\bibitem[{{Roberts} {et~al.}(1987{\natexlab{a}}){Roberts}, {Goldstein},
  {Klein}, \& {Matthaeus}}]{roberts1987JGRb}
{Roberts}, D.~A., {Goldstein}, M.~L., {Klein}, L.~W., \& {Matthaeus}, W.~H.
  1987{\natexlab{a}}, \jgr, 92, 12023, \dodoi{10.1029/JA092iA11p12023}

\bibitem[{{Roberts} {et~al.}(1987{\natexlab{b}}){Roberts}, {Klein},
  {Goldstein}, \& {Matthaeus}}]{roberts1987JGRa}
{Roberts}, D.~A., {Klein}, L.~W., {Goldstein}, M.~L., \& {Matthaeus}, W.~H.
  1987{\natexlab{b}}, \jgr, 92, 11021, \dodoi{10.1029/JA092iA10p11021}

\bibitem[{{Shoda} {et~al.}(2019){Shoda}, {Suzuki}, {Asgari-Targhi}, \&
  {Yokoyama}}]{ShodaEA2019ApJL}
{Shoda}, M., {Suzuki}, T.~K., {Asgari-Targhi}, M., \& {Yokoyama}, T. 2019,
  \apjl, 880, L2, \dodoi{10.3847/2041-8213/ab2b45}

\bibitem[{Smith {et~al.}(2006)Smith, Vasquez, \&
  Hamilton}]{smith2006interplanetary}
Smith, C.~W., Vasquez, B.~J., \& Hamilton, K. 2006, Journal of Geophysical
  Research: Space Physics, 111

\bibitem[{{Stone} {et~al.}(2020){Stone}, {Tomida}, {White}, \&
  {Felker}}]{StoneEA2020ApJS}
{Stone}, J.~M., {Tomida}, K., {White}, C.~J., \& {Felker}, K.~G. 2020, \apjs,
  249, 4, \dodoi{10.3847/1538-4365/ab929b}

\bibitem[{{Totten} {et~al.}(1995){Totten}, {Freeman}, \&
  {Arya}}]{TottenEA1995JGR}
{Totten}, T.~L., {Freeman}, J.~W., \& {Arya}, S. 1995, \jgr, 100, 13,
  \dodoi{10.1029/94JA02420}

\bibitem[{{Tu} \& {Marsch}(1994)}]{TuMarsch1994JGR}
{Tu}, C.~Y., \& {Marsch}, E. 1994, \jgr, 99, 21,481, \dodoi{10.1029/94JA00843}

\bibitem[{{Wang} {et~al.}(2023){Wang}, {Chhiber}, {Cuesta}, {Roy}, {Pecora}, \&
  {Matthaeus}}]{ChhiberEA2023inprep}
{Wang}, V., {Chhiber}, R., {Cuesta}, M.~E., {et~al.} 2023, in prep.

\bibitem[{{Yang} {et~al.}(2019){Yang}, {Li}, {Li}, {Zhang}, {He}, \&
  {Feng}}]{LPYangEA2019MNRAS}
{Yang}, L.~P., {Li}, H., {Li}, S.~T., {et~al.} 2019, \mnras, 488, 859,
  \dodoi{10.1093/mnras/stz1747}

\bibitem[{{Yang} {et~al.}(2017){Yang}, {Matthaeus}, {Shi}, {Wan}, \&
  {Chen}}]{YangEA2017PoF}
{Yang}, Y., {Matthaeus}, W.~H., {Shi}, Y., {Wan}, M., \& {Chen}, S. 2017,
  Physics of Fluids, 29, 035105, \dodoi{10.1063/1.4979068}

\bibitem[{{Yang} {et~al.}(2016){Yang}, {Shi}, {Wan}, {Matthaeus}, \&
  {Chen}}]{YangEA2016PhRvE}
{Yang}, Y., {Shi}, Y., {Wan}, M., {Matthaeus}, W.~H., \& {Chen}, S. 2016, \pre,
  93, 061102, \dodoi{10.1103/PhysRevE.93.061102}

\bibitem[{{Yang} {et~al.}(2021){Yang}, {Wan}, {Matthaeus}, \&
  {Chen}}]{YangEA2021JFM}
{Yang}, Y., {Wan}, M., {Matthaeus}, W.~H., \& {Chen}, S. 2021, Journal of Fluid
  Mechanics, 916, A4, \dodoi{10.1017/jfm.2021.199}

\bibitem[{{Zank} {et~al.}(2017){Zank}, {Adhikari}, {Hunana}, {Shiota}, {Bruno},
  \& {Telloni}}]{Zank2017ApJ835}
{Zank}, G.~P., {Adhikari}, L., {Hunana}, P., {et~al.} 2017, \apj, 835, 147,
  \dodoi{10.3847/1538-4357/835/2/147}

\bibitem[{{Zank} \& {Matthaeus}(1992)}]{zank1992JPP}
{Zank}, G.~P., \& {Matthaeus}, W.~H. 1992, Journal of Plasma Physics, 48, 85,
  \dodoi{10.1017/S002237780001638X}

\bibitem[{{Zank} \& {Matthaeus}(1993)}]{zank1993nearly}
---. 1993, Physics of Fluids, 5, 257, \dodoi{10.1063/1.858780}

\bibitem[{{Zank} {et~al.}(1990){Zank}, {Matthaeus}, \&
  {Klein}}]{ZankEA1990GeoRLDensity}
{Zank}, G.~P., {Matthaeus}, W.~H., \& {Klein}, L.~W. 1990, \grl, 17, 1239,
  \dodoi{10.1029/GL017i009p01239}

\bibitem[{{Zank} {et~al.}(2021){Zank}, {Zhao}, {Adhikari}, {Telloni}, {Kasper},
  \& {Bale}}]{zank2021PhPtransport}
{Zank}, G.~P., {Zhao}, L.~L., {Adhikari}, L., {et~al.} 2021, Physics of
  Plasmas, 28, 080501, \dodoi{10.1063/5.0055692}

\end{thebibliography}

\bibliographystyle{aasjournal}

% \listofchanges

\end{document}